\documentclass{egpubl}
\DeclareUnicodeCharacter{0301}{\'{e}}
\JournalSubmission    
\usepackage[T1]{fontenc}
\usepackage{dfadobe}  
\usepackage{mathptmx} 

\biberVersion
\BibtexOrBiblatex
\usepackage[backend=biber,bibstyle=EG,citestyle=alphabetic,backref=true]{biblatex} 
\electronicVersion
\PrintedOrElectronic

\ifpdf \usepackage[pdftex]{graphicx} \pdfcompresslevel=9
\else \usepackage[dvips]{graphicx} \fi

\usepackage{egweblnk} 
\usepackage{cleveref}

\title[MOTIV]{MOTIV: Visual Exploration of Moral Framing in Social Media}

\author[A. Wentzel et. al]
{\parbox{\textwidth}\centering A. Wentzel$^{1,2}$\orcid{0000-0002-2003-2750}, L. Levine$^{1}$, V. Dhariwal$^{1}$, Z. Fatemi$^{1}$, A.Bhattacharya$^{1}$\orcid{0000-0003-3647-3363}, B. Di Eugenio$^{1}$\orcid{0000-0003-1706-2577},  A. Rojecki$^{1}$\orcid{0000-0003-0773-9011}, E. Zheleva$^{1}$\orcid{0000-0001-7662-2568}, G.E. Marai$^{1}$\orcid{0000-0002-7212-9669}
\\
{\parbox{\textwidth}{\centering $^1$University of Illinois Chicago}}\\
{\parbox{\textwidth}{\centering $^2$Corresponding Author awentze2@uic.edu}}
}

\begin{document}
\teaser{
    \centering
    \includegraphics[width=.85\textwidth]{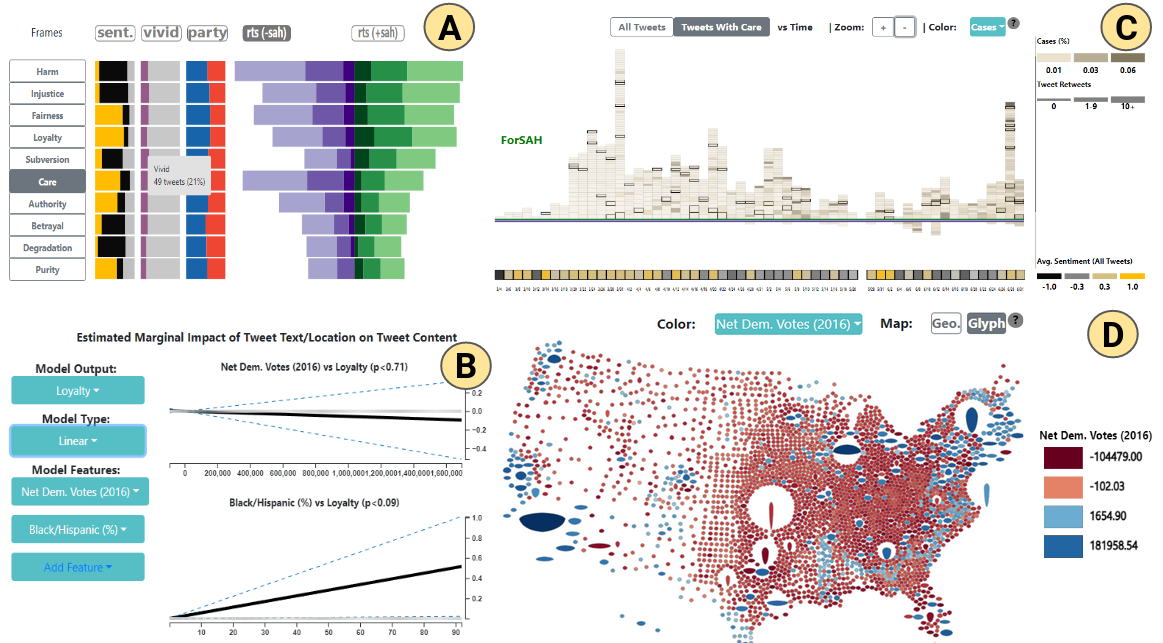}
    \caption{MOTIV visualization of moral framing on social media in 2020. (A) Summary panel showing tweet feature such as sentiment and tweets for or against the topic (B) Model building view showing inference scores for county votes withing each county vs tweets expressing Loyalty, derived from a generalized linear model (C) Timeline of tweets, along with retweet count, COVID-19 cases, and sentiment. (lower bar). Counties from LA are shown in bold (D) Glyph-based map of counties showing 2016 voting history (color), voting age population (width), and tweets (height).}
    \label{fig:teaser}
}

\maketitle
\begin{abstract}
We present a visual computing framework for analyzing moral rhetoric on social media around controversial topics. Using Moral Foundation Theory, we propose a methodology for deconstructing and visualizing the \textit{when}, \textit{where}, and \textit{who} behind each of these moral dimensions as expressed in microblog data. We characterize the design of this framework, developed in collaboration with experts from language processing, communications, and causal inference. Our approach integrates microblog data with multiple sources of geospatial and temporal data, and leverages unsupervised machine learning (generalized additive models) to support collaborative hypothesis discovery and testing. We implement this approach in a system named MOTIV. We illustrate this approach on two problems, one related to Stay-at-home policies during the COVID-19 pandemic, and the other related to the Black Lives Matter movement. Through detailed case studies and discussions with collaborators, we identify several insights discovered regarding the different drivers of moral sentiment in social media. Our results indicate that this visual approach supports rapid, collaborative hypothesis testing, and can help give insights into the underlying moral values behind controversial political issues.
\\Supplemental Material: \url{https://osf.io/ygkzn/?view_only=6310c0886938415391d977b8aae8b749}
\end{abstract}  
\section{Introduction}
Social media has become a center of discussion of heated political discourse, ranging from the response to local government policy, to the rise of the \#Me Too and \#Black Lives Matter protests in the US, and the shifting narratives that drove increasingly polarized reactions to the COVID-19 pandemic. The shift towards social media for discussing divisive political issues has made morality a vehicle for political messaging of all kinds, from social movements, misinformation and political propaganda through the use of modern moral panics~\cite{pepin2022president}. In addition, the effect of the pandemic has inspired a renewed interest in understanding driving factors in the propagation of ideas on social media~\cite{dagstuhl2022framing,dagstuhl2024reflections}. Analyses of social media discourse have attempted to either distill quantifiable text features that summarize popular topics~\cite{neri2012sentiment,twitinfo} or identify the content spread by major influencers such as news outlets~\cite{whisper}. However, such basic text features often miss key information about users’ motivations and personal values.

One approach that can help quantify users’ motivations when considering social media dynamics is Moral Foundations Theory (MFT)~\cite{graham2018moral}. Moral Foundation Theory is a psychological tool that proposes using a set of "Moral Foundations" as a basis to explain human reasoning. In this model different Moral Frames, such as Loyalty or Authority, can give insight into the nature of political discourse that is missing in traditional social media analysis approaches which consider only demographic and social factors. MFT has been applied to predicting social dynamics~\cite{dehghani2016purity, brady2017emotion}, reaction to violent protests~\cite{mooijman2018moralization,giersch2019punishing}, responses to hate speech~\cite{wilhelm2019gendered}, and reaction to appeals for charity~\cite{wang2021role,hoover2018moral}.

Social media analysis of ongoing topics gave several challenges. From a computational perspective, the short, informal nature of tweets, reliance on context and linked media in tweets, and the difficulty in understanding moral framing make traditional natural language processing (NLP) approaches such as semantic dictionaries and neural models relatively ineffective. For example, the sentence “Fauci said we should stay home!" could be a pro-SAH tweet that is expressing Authority by following an expert. On the other hand, this same sentence posted by a different individual may use this as an anti-SAH expression of Freedom, depending on the individuals' and their audience's feelings about Anthony Fauci. Therefore, meaningful analysis of social media data benefits from human-in-the-loop expert input. In turn, the strong link between location and personal values, and the large number of features at play mean that data visualization and interactive machine learning could be a valuable help in leveraging this expert input.

Additionally, since our proposed work included the rapid development of a dataset containing tweets annotated with Moral Frames, stance, and other features related to COVID-19, the nature of the data and resulting design goals changed rapidly as our understanding of the data updated. These changes ranged from the expected size of the dataset, to the features used, and the identify retweets and multiple tweets between users, which quickly changed what was feasible with the data. Furthermore, the rapid nature of our collaboration in conjunction with the changing nature of the pandemic exacerbated issues with collaborations. Our collaborators come from a variety of backgrounds, and thus had different baseline expectations and workflows, and the short nature of the project meant that our collaboration had limited time to mature. While the design process was a challenge, it did yield several domain specific insights that were published by our collaborators in addition to our visualization work~\cite{APSApaper,elenaDSAA}. Our design process had to meet significant challenges, from vague requirements to ongoing data foraging.

Visualization of MFT social media data poses several challenges. First, tweets may feature more than one MF, making succinct summarization difficult. Furthermore, social media discourse is a rich field which involves a large number of both textual and demographic features. In addition, because of the need to capture context in social media trends, the resulting data is large scale, and both temporal and geospatial. Last, the data is analyzed at multiple levels of detail, from high level trends in large corpuses to detailed content and local context. As a result, a solution needs to be able to handle a large number of different features while still maintaining an acceptable level of visual simplicity to make the system usable for clients with limited visual literacy.  

In response to these challenges, we present a novel integrated visual framework for analyzing Moral Frames in social media. This framework is designed in collaboration with domain experts in NLP, machine learning, communications, and social science. Our collaborators are keenly interested in analyzing how differences in messaging affect public sentiment regarding controversial issues related to public health and welfare, in order to improve public messaging for social good. Our contributions are: 1) An analysis of the activities and workflows needed for the Moral Frame analysis of discourse; 2) The activity-centered design and implementation of MOTIV (Media Opinion Trend Inference and Visualization), a visual analysis system for exploring annotated, geotagged social media MF data; 3) An evaluation with domain experts in multiple fields; and 4) Lessons learned from the design process, with particular emphasis on working with an evolving dataset and during the data foraging and data understanding phases, as well as challenges when working with domain experts with limited visual literacy and different design goals.

\section{Related Work and Background}

\noindent\textbf{Moral Foundation Theory} is a model for analyzing social dynamics by identifying underlying "moral frames" implicit in the values expressed by individuals within a group. MFT was introduced by Graham et al, as a way of discussing the difference in moral values among groups. Graham's model used 5 (later expanded to 6) foundations, which are each split into positive (virtue) and negative (vice) orientations. For example, one frame is Care/Harm. "Care" is the virtue that is defined as  "the need to help or protect oneself or others". "Harm" is the contrasting vice, which deals with "fear of damage or destruction to oneself or others"~\cite{graham2013moral}. The 6 pairs of 12 Moral Frames are: Care | Harm, Loyalty | Betrayal, Authority | Subversion, Purity | Degradation, Fairness | Injustice, and Freedom | Oppression. Of these Moral Frames, Loyalty, Authority, and Purity are often referred to as “binding frames”, which are the ones more strongly associated with conservatism. In contrast, Care and Fairness are  “individualizing frames”, which are associated with Liberalism. Freedom and Oppression are unique Moral Frames, proposed to better capture the viewpoints of Libertarians~\cite{iyer2012understanding}.

MFT has been used throughout social science in order to explore different values within groups, such as how individuals within different political parties may value different moral frames differently~\cite{koleva2012tracing}. Moral arguments affect individuals' stance~\cite{rezapour2021incorporating}. Moral foundations have also been tied to public health behaviors. For example, vaccine hesitancy is associated with Purity and Liberty, while pro-vaccine messaging focuses on Care and Harm~\cite{amin2017association}. Research has suggested that demographics influence moral framing, with women being more affected by Care/Harm, Injustice, and Purity~\cite{wilhelm2019gendered, atari2020sex}, and Authority, Loyalty, and Purity are linked to conservative viewpoints in White Americans, but not African Americans~\cite{davis2016moral}.

Alternatives to MFT include the theory of Moral Motives~\cite{janoff2013surveying}, Dyadic Morality~\cite{schein2018theory}, and Relationship Regulation Theory~\cite{rai2011moral}.
We use MF theory over these alternatives as it provides the most diverse set of moral dimensions, and has been successfully used in popular textual frame analysis models in political communication~\cite{entman1993freezing}. Research has also shown that moral judgment combined with emotion is a primary driver of viral spread in social networks and public health~\cite{valenzuela2017behavioral, heimbach2015content, wang2021moral}. MFT is thus a valuable tool for understanding how discourse develops around politically divisive issues on social media. In particular, we look at the MFT rhetoric surrounding politicized issues in the U.S, how moral valuation and stance relate to demographic factors, and the underlying Moral Foundations driving discussions on Twitter.

\noindent\textbf{Social Media Analysis and Visualization}
Many studies have been done regarding social media responses to different topics~\cite{jang2021tracking,chang2021people,chandrasekaran2020topics}. Such studies include responses to the COVID-19 pandemic~\cite{aiello2021epidemic}, feelings about public health policy such as vaccine mandates~\cite{doogan2020public,hu2021revealing}, and climate change~\cite{diakopoulos2014identifying}.

Chen et al.~\cite{chen2017social} provides an overview of common visualization goals: visual monitor, feature extraction, event detection, anomaly detection, predictive analysis, and situational awareness. Our system uniquely merges aspects of feature extraction, event and anomaly detection and stance detections, with the integration of moral foundation theory and enriched demographic features for applications in politics and journalism.

Several visualization systems have looked at how information spreads within communities on social media. Google+ Ripples~\cite{googleripples} shows communities of google+ users using Euler diagrams~\cite{rodgers2008general}. Visualization of social network information has also been used in journalism to visualize news coverage~\cite{mishra2022news}, and identify misinformation~\cite{dant2011behind,karduni2019vulnerable}.

Several systems have been built for real-time detections, such as visualizing information spread on social media using retweets and topic sentiment~\cite{whisper,rmap}, and real-time topic clustering~\cite{knittel2021real}, but don't incorporate geospatial information. In contrast, several systems have used integrated maps and text summarization for event detection~\cite{leadline} and disaster response~\cite{scatterblogs2}. However, these systems focus only on social media data, and do not analyze moral frames or augment their data for more detailed analysis beyond simple sentiment.

For temporal analysis, other systems have focused on temporal progression of topics using sentence trees~\cite{sententree}, timelines~\cite{opinionflow} and custom encodings~\cite{digitalbackchannels}. Other works have explicitly focused on polarized topics~\cite{cobridges} and stance detection~\cite{kucher2020stancevis}.
Some work has integrated human-machine mixed analytics to detect "anomalous" threads~\cite{fluxflow} and bots~\cite{targetvue}, which both rely on a mixture of timeline visualization and glyphs. However, none of these systems tie their discourse to demographics or MFT framing.

Other methods have linked spatial and temporal information through methods such as Spatio-temporal clustering~\cite{mobilitygraphs} and flow maps~\cite{kim2018data}. Other systems use linked views for monitoring events on Twitter~\cite{twitinfo}, and journal articles~\cite{park2016supporting}. However, no existing systems have incorporated other spatial information such as demographics or regional political ideologies.

\section{Design}

\subsection{Design Process}
MOTIV was developed via remote collaboration between four different research groups between May 2020 and February 2022 as the result of a RAPID~\cite{NSFcall} grant intended to fund projects to help inform and educate the public about COVID-19 safety measures. MOTIV was designed alongside the development of our dataset and required rapid updates to our design requirements and goals. Our design process is based on an Activity Centered Design (ACD) domain characterization process~\cite{marai2017activity}, which we modified as a result of our irregular program circumstances. 

The core group consisted of two researchers in communications, three NLP researchers, two researchers in causal inference in social media, and two visual computing researchers, all of whom are listed as co-authors. The team met remotely twice a month to discuss updates, identify project goals, discuss progress in data analysis, augment results from analyzing and annotating tweets, and produce visual representations based on data gathered from various sources. 

At the beginning of the project, we used interviews and notes taken during meetings to develop the task analysis and project requirements, which were updated regularly at meetings as we explored potential data sources. Since our project was based on an emerging topic at the time (Stay-at-Home orders), we frequently worked on high-fidelity or functional prototypes of datasets as they were being developed and shared during group meetings. Results from these sessions guided future directions for the project and future datasets, and feedback informed updates to existing design requirements. Due to the nature of the data and collaboration, we focused on developing encodings that show as much data as possible and then refining them into simpler encodings that were more accessible to collaborators with moderate visual literacy (see supplement). Because the project aimed to support domain experts, not novice users, our design process focused on identifying existing workflows and activities performed by the domain experts, and building solutions that extend these activities.

\subsection{Activity and Task Analysis}
\begin{figure*}[htp]
    \centering
    \includegraphics[width=.9\textwidth]{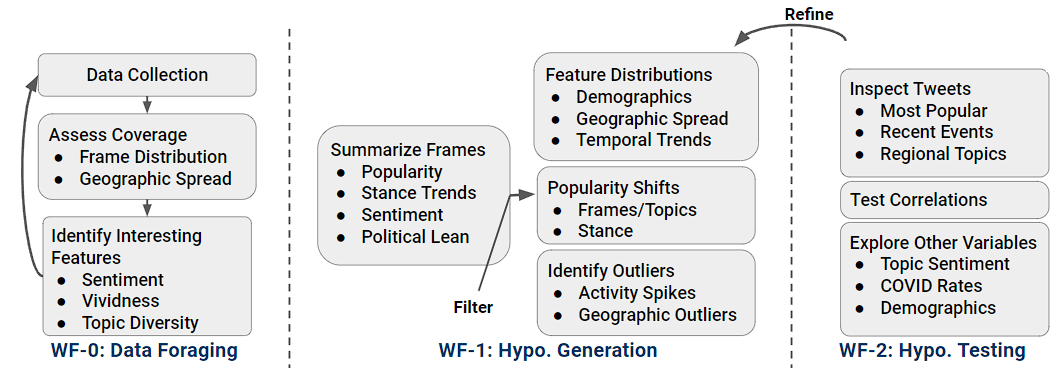}
    \caption{Workflows: (WF 0) Data foraging, where data is iteratively collected and analyzed to identify the quality of coverage and interesting features. (WF 1) Hypothesis generation, where moral frames are analyzed to identify interesting findings. A summary view is used to identify interesting frames which are filtered and assessed in more detail. (WF 2) Hypothesis testing, where observations in (WF 1) are confirmed by drill-down or correlation testing. Insights are used to guide future investigations in (WF 1).}
    \label{fig:workflows}
\end{figure*}

MOTIV was designed alongside collaborators in Communications and NLP, with an emphasis on supporting the Communications users in the final version of the interface, while earlier prototypes were intended to support NLP research in the development of a moral-frame annotated dataset. 

In the beginning of this project, our aims were to support the development of a usable dataset along with collaborators in NLP and causal inference, which largely modeled the \textit{foraging} loop in the sensemaking process~\cite{pirolli2005sensemaking}. Specifically, our collaborators, both computing experts and communication scientists, were interested in ways of generating a relevant tweet corpus, and assessing its geopolitical and temporal coverage.

Given this characterization, we found that our collaborators were interested in multiple, interrelated workflows (\cref{fig:workflows}).
\textit{Foraging} (WF-0), is where all researchers assess the quality of the tweets, the coverage of the dataset in terms of moral foundations, time, location, and stance, and the distribution of potentially relevant features such as sentiment or vividness.
\textit{Hypothesis generation} (WF-1), is where researchers searched for major trends within the social media data, such as a general increase in the tweets about Liberty, and then developed hypothesis around potential causes of these trends, such as these Liberty tweets being driven by people from rural areas. Finally, during the \textit{Hypothesis testing} phase (WF-2), researchers looked for ways to verify the causes of these trends, such as by looking at the correlation between population and Liberty tweets or investigating the events that co-occur with a spike in pro-Liberty tweets. The findings from the second stage would then feed back into WF-1.

During the data analysis state, we found that our interdisciplinary team's main interests ranged from examining how different socioeconomic and demographic factors relate to stance and moral framing with respect to controversial issues, as well as what textual factors such as "vividness" and "sentiment" affect tweet popularity. We found that our collaborators tended to model macro-level social dynamics as a feedback system, in which overall trends tended to be guided by two phenomena of interest: 1) grassroots memetic propagation of ideas in response to larger social movements, and 2) disruption events when a notable story or individual causes a shift in the online zeitgeist. Their research activities are thus focused on identifying and explaining these types of phenomena and how Moral framing factors into them. In this way, MF serves as a lens to describe larger trends within social movements, while also serving as a reflection of how disruptive movements are viewed by others. We focused on identifying tasks that could not be done by individual researchers via their standard workflows:

\begin{itemize}
	\item \vspace{0em} A1. \textit{Summarize relationships between Moral Frames and demographic and political factors:} When investigating Moral Foundations on Twitter, our collaborators started with investigating key features and trends surrounding each Moral Frame (WF-0). They were interested in how political affiliation, tweet content, and popularity differed between each Moral Frame, and how this affected tweet stance and virality. Collaborators focused on investigating high-level relationships in the data using summarization before determining which MF to explore in detail (WF-1).
	\item \vspace{0em} A2. \textit{Understand temporal trends:}
Beyond high-level relationships, our team was very interested in exploring temporal trends in the popularity of each MF, with a focus on points when a topic would drastically change in popularity (WF-1), which could then be tied back to inciting events (WF-2). Our team was interested in the effect that changes in COVID-19 cases and lockdown orders had on MF trends, so a major requirement is the ability to include details of case rates alongside tweet popularity.
	\item \vspace{0em} A3. \textit{Identify characteristics and Moral Frames of viral tweets:} Polarized discussion can be strongly affected by a small number of particularly viral ideas. Our collaborators were interested in identifying the most viral tweets, and their underlying Moral Frames. Identifying important tweets can help identify events or tweets that drive changes in temporal trends (WF-1). Additionally, identifying commonalities within viral tweets provides insights into potentially interesting features (WF-0), and how Moral Framing is viewed by different groups (WF-2).
	\item \vspace{0em} A4. \textit{Understand the geographic distribution of each Frame within social context:} Moral framing is heavily tied to political ideology and culture in literature, and regional differences are thus a major factor in how Moral Frames propagate within different groups. As a result, we wished to identify the geographical distribution of our tweets with different moral frames (WF-0), as well as overlap with factors such as income, political leanings, and COVID-19 cases (WF-1), with a focus on how Moral Frames vary based on the socioeconomic factors in the local area in response to state and country-wide mandates (WF-2).
	\item \vspace{0em} A5. \textit{Verify hypotheses about meaningful relationships in the data:}  Once hypotheses and potential relationships in the data were identified (WF-1), our collaborators often resorted to performing statistical testing to verify these findings. This was valuable for identifying features that would be useful for future models for our NLP researchers (WF-0), and validating findings from our communications researchers (WF-2). Thus, MOTIV needed to be able to identify causal effects while accounting for confounders, in a way that was immediately available to both analysts and non-analysts during sessions.
\end{itemize}

Non-functional requirements included online availability for remote collaborators. Because the datasets required expert annotations, our design needed to be usable with up to around 2,000 tweets, with potential to scale to larger datasets if better automated methods become available.

\subsection{Data and Architecture}
\begin{figure}[h]
    \centering
    \includegraphics[width=\linewidth]{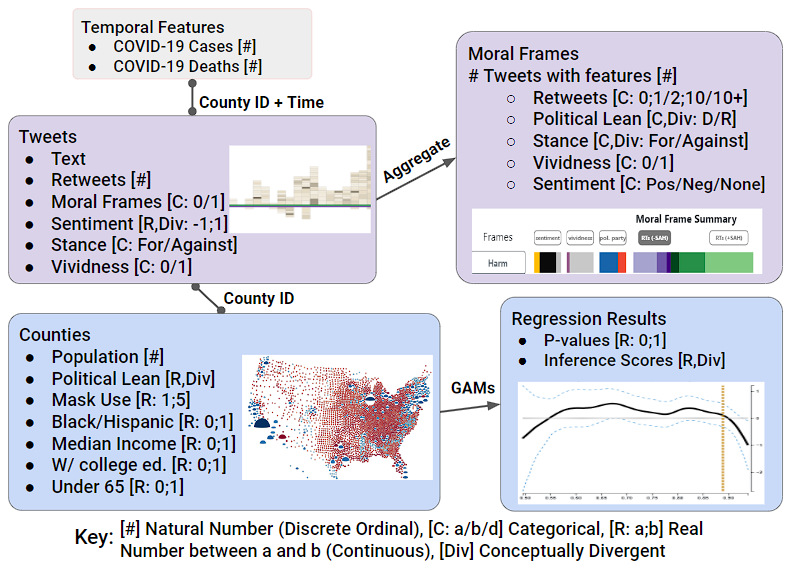}
    \caption{Data Abstraction. Tweets are labeled with textual features, and augmented with county-level data using the timestamp and geolocation data. Tweets are aggregated by MF and county for summarization. Generalized Regression Models model county demographics and aggregated tweet statistics to generate partial dependence plots in the inference views. County FIPS code is used to link all sections of the interface during brushing.}
    \label{fig:abstraction}
\end{figure}

MOTIV was originally designed around a dataset of stay-at-home (SAH) tweets during the beginning of the COVID-19 pandemic. Our dataset was gradually constructed and updated during the foraging stage of the project, with multiple data sources being added in gradually. Our twitter corpus of geotagged annotated tweets was constructed as described by Fatemi et al.~\cite{elenaDSAA}, which we briefly summarize here. First, a sample of 87M tweets were taken from March 1 to June 30, 2020, from a dataset of COVID-19-related tweets~\cite{chen2020tracking}. Through an interactive process with our collaborators, we identified a set of potentially relevant tweets that contained the following keywords: home, open, quarantine, inside, and lockdown. These tweets were then hand-annotated by Moral Frame experts with the following information: 1) stance - if the tweet was in support of or against SAH, 2) whether the tweet contained specific or vivid descriptions (vividness), and 3) which of the 12 Moral Frames were expressed in the tweet, if any.

To reduce burden on our manual annotators, sentiment score was annotated using the sentiment analysis tool Vader~\cite{hutto2014vader}, which uses a rule-based lexical system to determine whether the content in the tweet expressed positive emotions (e.g happiness) or negative emotions (e.g anger), without requiring training data. VADER has been shown to outperform other baselines, including human annotators, for sentiment analysis~\cite{ribeiro2016sentibench}. Scores of $> 0.25$ were defined as "positive", scores of $< -0.25$ were defined as "negative" sentiment, and middling scores were defined as "neutral".

We mapped the geolocation associated with each tweet to each of the 3113 US counties, excluding Antarctica, as follows. First, we obtained the bounding box of each geotagged tweet from the Twitter metadata. We then calculated the area of overlap between each bounding box and the borders for each county. Each tweet was assigned to the county with the highest percentage of overlap, and tweets that did not have at least 25\% overlap with a single county were excluded. After removing tweets that did not take a stance, or did not express a Moral Frame, we ended up with 1483 geotagged tweets from the US that were determined to be relevant to SAH orders.  

Overall, our system considers two data items: tweets, and counties, which are connected via geolocation. For tweets we use the geotag to identify features taken from the corresponding count. On the county level, we incorporate 2018 census data~\cite{medsl2018elections,census2018},  voting ratings for each county from 2018~\cite{medsl2018elections}, a self-rated mask usage survey from the new york times~\cite{nyt2020mask}, and COVID-19 cases and death rates for the time period covered by the dataset~\cite{johnshopkins2020dashboard}.

Political leaning is encoded as the number of votes for the democratic party minus the votes for the republican party in the 2016 presidential election, based on collaborator input.
We also aggregate the total number of tweets with each moral frame each stance within each county, which results in 14 different continuous values for each county.
Additional attributes selected during the foraging stage, and the data abstraction are detailed in Fig.~\ref{fig:abstraction}.

Later on, MOTIV was further used to analyze a second dataset taken from  the Moral Foundations Twitter corpus~\cite{hoover2020moral} to compare geotagged tweets associated with the \#BlackLivesMatter (BLM) movement between 2014 and 2016. We encoded stance using hashtags: tweets that contain more hashtags for BLM to be in support, while tweets that contain more hashtags related to the All or Blue Lives Matter (ALM) to be opposed. Tweets that contained an equal number of hashtags for each side were excluded as we could not be confident in their stance.
In total, we identified 1051 tweets in support of the BLM movement and 854 tweets in support of the ALM movement.

Data processing is implemented in python using the Flask and Pandas packages. The front-end is implemented as a web app using JavaScript with the d3.js and React libraries. Generalized additive models were implemented using the pyGam package.

\subsection{Layout Design}
To support the five main activities (A1-A5), and both foraging and hypothesis-related workflows, MOTIV uses multiple coordinated views which were developed gradually as the dataset was being developed. The four panels support each one main activity, whereas the view coordination supports insights into multiple dimensions of the data. The entrance to our interface is a Moral Frame overview panel that shows sentiment, political party, stance and retweets aggregated by each frame. Once a frame of interest is identified, the analyst selects that frame, which loads detailed views in the other panels (A1). To get an overview of temporal trends alongside COVID rates, we include a novel timeline view (A2, A3). To view geospatial trends, we use a novel glyph-based map that encodes demographics, MF popularity, and population for each county in the US (A5). Finally, an Inference panel allows for building predictive models and visualizes their partial dependence curves, which helps identify the relationship between individual features and demographics (A5).

We designed encodings to help capture foraging and hypothesis supporting patterns and outliers (A1-A4). To deal with the issue of “misleading” patterns, we then introduced an inference panel and tooltip details, to be used to validate findings with greater fidelity (A5).

MOTIV was designed to support our collaborators’ specific research needs, as opposed to novice users, and so our novel encodings benefit from participatory design and from visual scaffolding~\cite{marai2015visual}. Still, as our domain experts wished to be able to share the system with novice researchers in their groups, MOTIV also provides explicit legends and visual explanations on demand for custom encodings.

\subsection{Summarization Panel}
The summary view shows distributions of tweet-features that expressed a given Moral Frame. We use rotated stacked bar-charts to encode tweet features such as stance and vividness. The stance stacked bar charts are further broken up by the number of retweets, to indicate the overall popularity of each Moral Frame. Bars are aligned horizontally with Moral Frames, to allow for a side-by-side comparison of part-to-whole relationships. The panel also supports sorting the order of frames based on each feature, to better show frames with the highest or lowest incidence of a specific feature value. The linked panels will update to filter by tweets with the selected Moral Frame. 

Earlier design iterations included variants of parallel coordinate plots and correlation matrices. However, these were deemed to be unnecessary complex by our collaborators. 

\subsection{Timeline Panel}
Visualizing temporal information is important for understanding how discourse evolves over time, how public sentiment evolves in response to major events, and how these events give context to unexpected patterns in the data. To support this type of analysis, we use a novel Timeline panel that encodes each tweet, as well as the tweet's date, popularity, stance, and geolocation data (such as COVID-19 rates) over time. Our timeline supports inspection at two levels of granularity: overall trends and major events, and the context of events by visualizing the details of popular tweets. 

\begin{figure}[htbp]
    \centering
    \includegraphics[width=\linewidth]{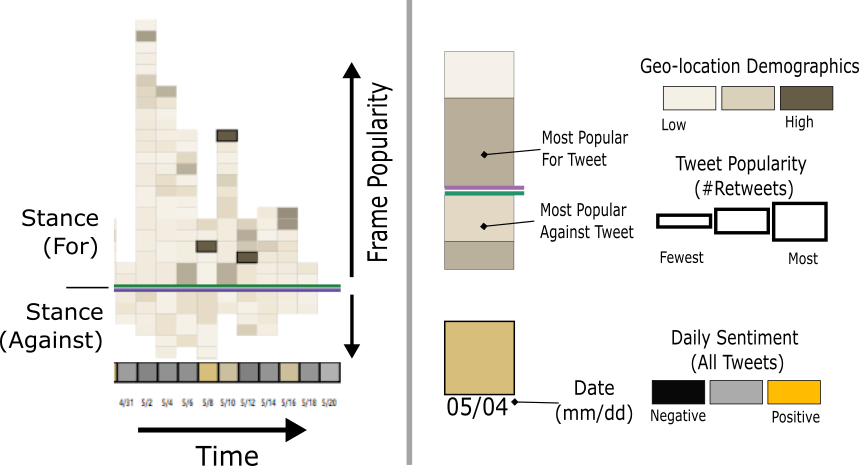}
    \caption{Outline of the timeline encoding. (Left) Timeline over a period of 10 time bins. Individual tiles encode tweets within the time bin. Tweets in for SAH are above the central axis while tweets against SAH are below the axis. Total tile height serves as a proxy for overall popularity. (Right) 4-tweet example encoding for a single time bin, annotated with the date in mm/dd format (05/04). Tiles correspond to tweets, height encodes retweet count, and color encodes demographic information about the tweet's original location. Tweets for the topic are placed above the center line, whereas tweets against the topic are placed below. Tweets are placed such that the most popular tweets are closest to the x-axis. A square tile in the bottom timeline shows the tweet date where color encodes sentiment score across all tweets for that date.}
    \label{fig:timeline}
\end{figure}

In the layout, the X-axis is mapped to time, and aggregated into "bins", or windows of time, depending on the total number of dates covered in the dataset. Individual tweets within each time window are encoded as tiles, which are stacked within each window of time (\cref{fig:timeline}). Thus, each tile is positioned along the X-axis of the timeline according to the tweet date. To capture differences in tweet stance, the center of the chart is bisected horizontally along the X-axis. Tweets that are in support of a topic (e.g., for SAH) are positioned above the center axis, whereas tweets opposed to that topic  are placed below the line, to allow direct stance comparison.

To support tweet popularity analysis, we sort the tweets along the Y-axis based on their popularity, such that the most popular tweets are always close to the center axis of the timeline. Because our goal is to capture the \textit{popularity} of each Moral Frame, and not simply the number of tweets, the height of each tweet is scaled
according to the number of retweets, such that more popular tweets contribute more to the overall height of the timeline.

\begin{figure}
    \includegraphics[width=\linewidth]{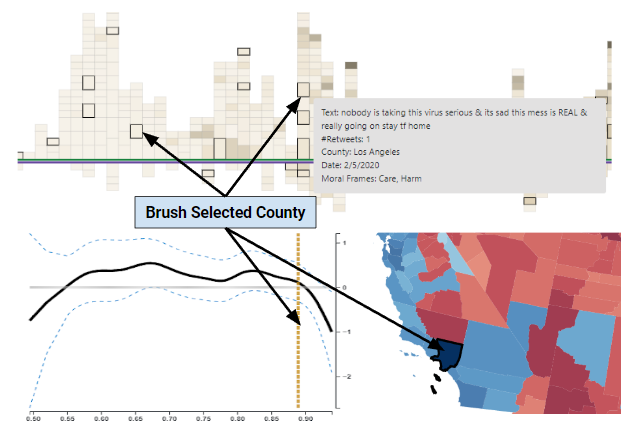}
    \caption{Examples of brushing and tooltip interactions in MOTIV: items associated with a selected county will be highlighted in the Map, Timeline, and Inference views; and the map and timeline views show additional details for counties or tweets via a tooltip when the user hovers the mouse over an item.}
    \label{fig:interactions}
\end{figure}

Finally, each tile is color-coded based on a user-selected demographic or tweet-specific features of interest. Details including text, location, and the Moral Frames expressed are provided via tooltip interaction. One can also filter the timeline to show only tweets expressing a certain Moral Frame. Selecting a tweet will highlight in the chart all tweets from the same county, as well as highlight the county the selected tweet is from, in the other panels (\cref{fig:interactions}).

We arrived at this custom encoding after exploring several popular variants of sparklines and steam graphs using a larger set of tweets without geotags, where color encoded sentiment over time. While this approach helped identify high-level trends, it also prevented inspection of individual tweets, which is important for understanding the context behind spikes in tweets. Additionally, since some features like COVID-19 cases were dependent on both location and time, it was important to map case rates to individual tweets in the timeline without aggregation.

\subsection{Geospatial Map Panel}
A major task was understanding how cultural and socio-econonomic factors influenced the spread of moral frames within different regions. To accomplish this, we include a county-level map showing local demographics and distributions of tweets with each moral frame. We experimented with choropleth maps, glyph-based encodings, and hybrids between the two. Because of the number of multiple variables to encode, designing this panel was particularly challenging. 

Our initial designs centered around choropleths, which were familiar to our collaborators, for showing tweets and demographics simultaneously. We experimented with using a mixture of color blending~\cite{hagh2007weaving}, texture blending~\cite{hagh2007weaving}, and overlayed glyphs (circles) to represent multiple variables, as recommended by Ware et al.~\cite{ware2020designing}. However, we found that it was still difficult to discriminate details around cities with high populations and small county area, which were regions of interest.

\begin{figure}[htbp]
\centering
    \includegraphics[width=0.8\linewidth]{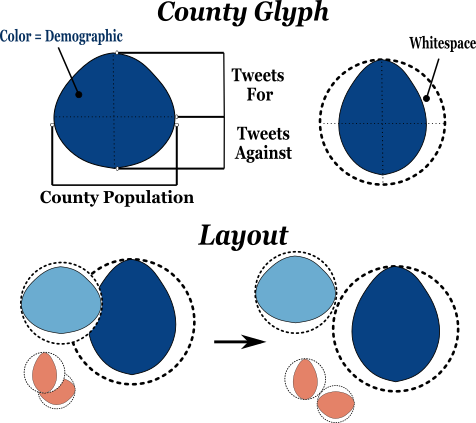}
    \caption{Glyph county map. (Top) Shape encoding for each county: width encodes population, while the upper and lower radius encode tweets for an against the topic of interest that express a certain MF. Color encodes a user-defined variable, which is voting history in the example. (Bottom) Glyph layout: Glyph positions start in the county center based on a Mercator projection, and are moved to avoid overlap using a force-directed layout that preserves whitespace in the glyph.}
    \label{fig:geomap}
\end{figure}

We eventually settled on a custom glyph-based map (\cref{fig:geomap}). The custom glyph uses a solid color to encode demographics, while shape encodes tweet density and population. Each glyph is drawn as a distorted ellipse, whose width encodes county population, whereas the upper radius encodes tweets within the Moral Frame in support of the topic, and the lower radius encodes moral tweets opposed to the topic. The resulting glyph is similar to a star-chart with a variable radius. We chose to use an ellipse over diamond shapes through experimentation, as we found that differences in the exaggerated degree of convexity of the curves of outlier counties served as a better pre-attentive cue than glyphs that use straight edges. A force-directed layout is used to adjust the position to prevent overlap between counties. This layout and the white space generated by the glyph helps emphasize counties with uneven tweet/population ratios. By comparing the size and shape of the glyph, one can easily identify and examine both major cities, and areas with a disproportionately high number of tweets with a given stance.

\subsection{GAM Inference Panel}
One major design goal during hypothesis generation and testing to confirm visual analysis inferences via statistical tests (A5) when determining what factors influence moral stance and tweet popularity. This activity faces two major issues. First, many demographic factors, such as population and COVID-19 rates can serve as confounding variables.
Second, the search space of potential confounders is too large to visualize all at once. To address these issues, we implemented an Inference panel that allows for interactive hypothesis testing. 

Our Inference panel is centered around the use of generalized additive models (GAMS)~\cite{hastie1990generalized}. GAMs are a class of predictive models~\cite{eilers1996flexible,wood2006generalized,larsen2015gam}.
that treat the predicted variable as the sum of individual functions of input variables, 


allowing users to visualize the relationship between each variable, while accounting for relationships between correlated variables that are taken into account in the multivariate model.

Our implementation consists of a control panel for interactively building a predictive model, and the partial dependence plots of each input variable 
(\cref{fig:teaser}-B). The control panel allows for the selection of the dependent variable being predicted, the input variables, and the type of shape function used in training the GAM. We included as potential predictors tweet-level features, such as the presence of a Moral Frame, or the number of retweets. Demographic factors, COVID-19 rates, and tweet content are included as potential input variables. 

The model allows for the use of either a spline or linear fit of the model. Spline curves allow for better representation of the distribution of the data, while linear models allow for more accurate reporting of p-values to identify statistically significant relationships in the data. 

The choice to use GAMs and partial-dependence-plots was decided after many design iterations. Early in the project, we used clustering with user-defined demographic or textual features to automatically generate intersectional groups that could be displayed as a series of bar charts, star charts, or modified sankey-diagrams. However, collaborators felt that the implementation was too complex to interpret quickly when performing hypothesis testing. In contrast, we found that GAMs and partial dependence plots were more grounded in the existing knowledge of collaborators who used regression and line-graphs regularly in their research, while showing fewer features at one time to reduce cognitive load to users.

\section{Evaluation}
MOTIV has been adopted as a research tool by our collaborators in communications, NLP, and causal inference, with the intention of supporting insights into SAH policy application in the U.S. For this reason, we demonstrate its capabilities in two case studies, reported here in abbreviated form, which were performed over several months by our collaborators and later used in publications in our collaborators respective fields~\cite{APSApaper,elenaDSAA}. In addition, we provide feedback from the target users.

Due to pandemic and work-from-home measures, the studies were completed remotely using the think-aloud technique with note-taking. We denote where these case studies correspond to activity workflows with the notation [WF]. Video summaries of the case studies are provided in the supplementary materials.

\subsection{Stay-at-home Attitudes and Dominant Moral Frames}
\begin{figure}[ht]
\begin{center}
    \includegraphics[width=0.9\linewidth]{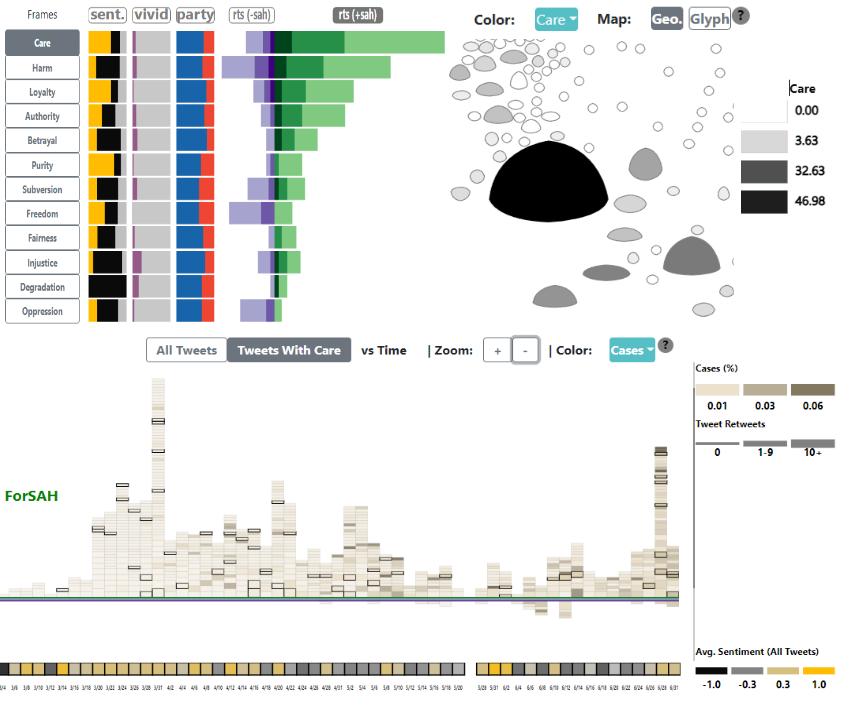}
    \caption{Overview for case study 1. (Top-Left) Summary panel of tweets in the SAH dataset, sorted by Popularity. Care and Harm are dominant, as all frames besides Freedom and Oppression are mostly for-SAH. (Top-Right) Glyph map of care-tweets by county, focused on L.A. (Bottom) Timeline of tweets expressing Harm. Major peaks occur at the end of March and June, with a smaller peak in April.}
    \label{fig:case1}
    \end{center}
\end{figure}

This case study focused on the creation of an annotated MF corpus and the subsequent analysis of Moral Frames as expressed in microblog data related to Stay at Home (SAH) orders in the U.S. (\cref{fig:case1}). 

Our collaborators were interested in which frames were dominant in the microblog data, as well as their vividness, popularity, sentiment, what temporal trends they followed, and the surrounding socioeconomic context around the tweets expressing each frame [WF 1]. Using the Summarization and Inference panels, the team confirmed relatively low popularity and a general lack of vividness across the corpus ($<$15\% vividness). The MF with highest vividness was Injustice (6 vivid tweets out of 25). Although the team had hypothesized a correlation between vividness and popularity, the Inference panel indicated a non-significant positive correlation (p $>$ .5). 

By sorting the most popular frames [WF 1], it became apparent that Care and Harm are the most popular frames expressed in Stay at Home tweets, and that they are both, surprisingly, predominantly in support of SAH orders. The communications experts noted that Care and Harm are complementary frames that form the virtue and vice around a single Moral Foundation, respectively, so this finding was intriguing. The group then noted that all “virtues” such as Care were correlated with higher sentiment (yellow in the sentiment column) than all “vices" (black in the sentiment column), such as Harm [WF 1].

The group was then extremely surprised to note that, aside from Freedom and Oppression, most other frames were also in support of SAH orders (Purple bars showing for tweets in the frame overview were larger than Green bars in moral frame overview) [WF 1]. These other frames were being expressed predominantly in democratic counties—even frames typically associated with conservative views, like Loyalty and Betrayal. Upon inspecting the timeline view, the group was able to confirm that most tweets are in support of SAH (predominantly above the centerline), and most tweets have low popularity (short tiles). In addition, they noted a correlation with increasing COVID-19 case numbers (darker tile shade), and overall more negative sentiment (more gray and black in the sentiment bar) as the pandemic evolves. By further examining individual tweets, they were able to determine that some viral tweets (taller tiles near the centerline) were, as expected, also vivid (e.g., \emph{"Protesters attacking governors for stay at home orders. Claim it infringes upon their rights. Know what else infringes upon your rights? DEATH."}). Several other popular tweets reflected counter-intuitive information (e.g., the news that most of the NYC new COVID-19 cases were people following SAH orders), influencer SAH tweets, or, again, vivid pleas from overwhelmed nurses and doctors working in intensive care units [WF 1]. 

A visual computing researcher then noticed in the Timeline panel several spikes in the number of SAH tweets on March 31st, May 2nd, and July 28th, and a significant and surprising drop around May 28th [WF 1]. This sparked a vivid discussion involving the county map. Communications experts inferred the peaks corresponded to the beginning and end of several regional lockdowns, whereas the drop corresponded to the onset of social unrest related to the George Floyd events and Black Lives Matter (BLM) movement in the US [WF 1]. 

Based on the same Timeline panel, the group noticed the first wave of anti-quarantine (below the center x-axis) tweets [WF 1], which, upon inspection in the Geospatial panel, appear to originate in counties with lower COVID-19 rates [WF 1]. Brushing the area around Los Angeles in the county map, we noticed suburban counties (small counties surrounding large glyphs representing cities) had a higher Harm/Care tweet ratio (short, dark glyphs) [WF 1]. The most senior communications expert hypothesized that tweets about Care originate mostly from large cities, whereas Harm is more evenly distributed among different suburban or rural populations [WF 1]. The group tested this hypothesis in the inference plot by showing the relationship between population and each frame in the Inference panel [WF 2]. Comparing both frames, the group found that Harm is indeed more prevalent in lower-population counties than Care (flat slope and smaller p-value) [WF 2]. 

The group concluded that the microblog discourse was generally in favor of SAH orders, with increasing negative sentiment as pandemic fatigue set in. Although Care was predominant, most of the other frames expressed were also overall in support of SAH, with several interesting anomalies. They also noted the data was biased towards urban areas (large, tall glyphs in the geospatial map). Near the end of May, the BLM rhetoric had nearly supplanted the SAH discourse, despite pandemic fatigue and an expectation of increasing conservative views. The team concluded that the public policy messaging which had targeted Care-for-others had been overall effective [WF 2].

\subsection{Moral Frames and Black Lives Matter}

\begin{figure}[ht]
\begin{center}
    \includegraphics[width=\linewidth]{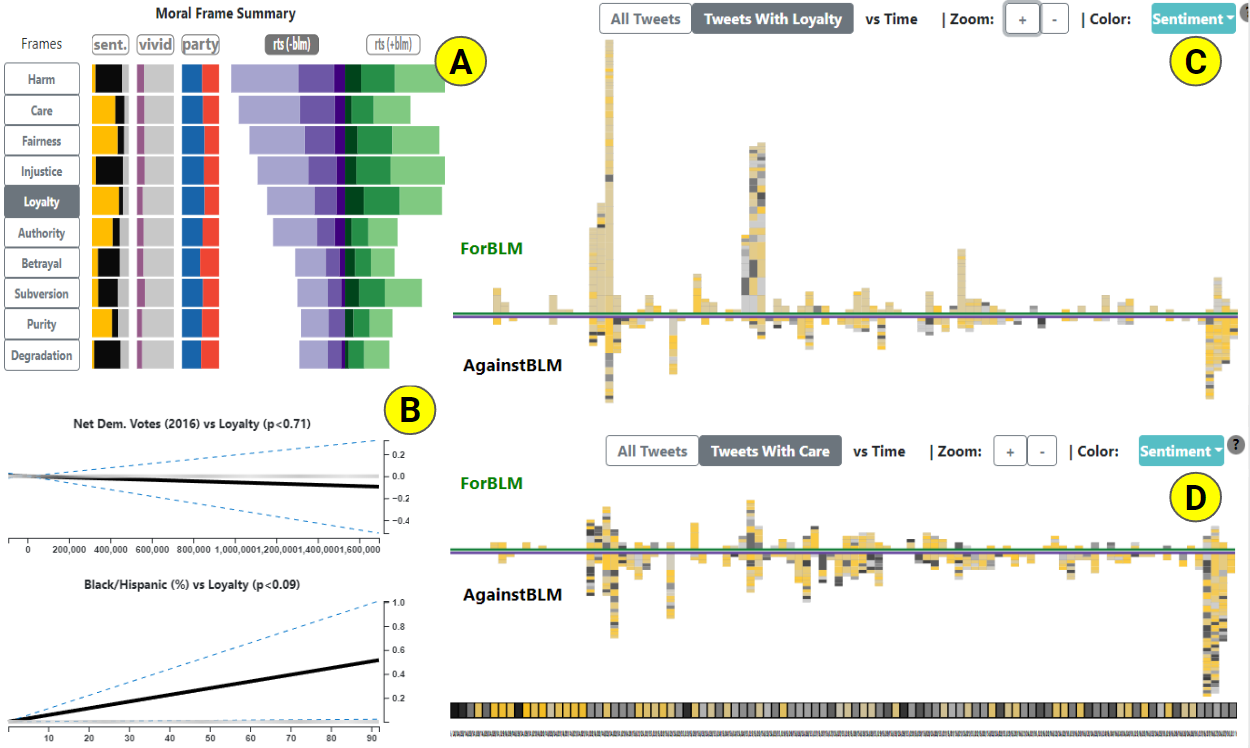}
    \caption{Overview of our BLM MF analysis. (A) Moral frame summary of tweets in the BLM dataset, sorted by percentage of tweets from democratic areas. Loyalty and Fairness are the dominant democratic frames, while betrayal is the most republican frame. (B) Correlations between demographics and frames.  Republican votes are correlated with tweets for Authority, while the percentage of Black or Hispanic individuals is the strongest predictor of pro-Loyalty tweets. (C) Tweet timeline of pro-Loyalty tweets colored by tweet sentiment. Spikes in \#BLM tweets occur around major protests. (D) Tweet timeline of pro-care tweets. A large spike in \#bluelivesmatter tweets occurs during July 2016, in response to a police shooting in Dallas.}
    \label{fig:blm}
    \end{center}
\end{figure}

This second case study uses a subset of the Moral Foundations Twitter corpus~\cite{hoover2020moral} to compare tweets associated with the \#BlackLivesMatter (BLM) movement and the \#AllLivesMatter (ALM) movement between 2014 and 2016. The \#BlackLivesMatter movement is a social movement that gained widespread popularity in 2014 in response to the disproportionate violence against African Americans, particularly by the police. The \#AllLivesMatter movement, among other movements, arose as a critical response to the BLM movement. Both movements have become central to political discussions in the United States around issues such as police protections and criminal justice reforms, and played a role in the 2016 US presidential election~\cite{eligon2015one}. Understanding the Moral Framework behind both movements can give insight into the driving forces behind these political movements.

The team started the  investigation with the Frame Summary Panel (\cref{fig:blm}-A) and sorting Moral Frames by political party [WF 1]. The frames most strongly associated with democratic areas (blue in the “party” column, top of the list) were Loyalty, Fairness, and Injustice. In contrast, Betrayal and Degradation were most often associated with more negative sentiment (black in the sentiment column) and republican areas (red in the “party” column, bottom of list) [WF 1].  

The team also noted that despite being relatively balanced politically, a majority of tweets that express Care are in support of ALM (purple column larger than green column), which is unexpected, given that prior literature suggests that Care is more strongly associated with political liberals, as is the BLM movement.
A communications researcher mentioned that Loyalty would be correlated with pro-BLM tweets since it is a “Binding Frame”, and decided to explore further by viewing pro-Loyalty tweets in the Timeline panel (\cref{fig:blm}-C) [WF 1]. Four major spikes in activity can be seen, 3 of which are predominantly for BLM (more tweets above the center axis) and from relatively democratic areas (blue rectangles), while one is for ALM with a higher percentage of Republican areas (red rectangles). Investigating the popular tweets from these time periods revealed the context behind these tweets: they are all tweets expressing solidarity for major protests related to police brutality: The Ferguson Protests~\cite{rothstein2015making}, the 2015 Baltimore Protests~\cite{mooijman2018moralization}, the 2015 Mizzou Protests~\cite{trachtenberg2018the}, and the 2016 Dallas Protests in which 5 police officers were murdered~\cite{maben2019re} [WF 2].  


Given the association between Care, political liberals, and SAH attitudes in our prior case studies, one researcher expressed interest in the fact that Care was not related to pro-BLM tweets “Care shows up in Republican areas, that’s strange”. In the timeline (\cref{fig:blm}-D), we see small spikes in activity around the Ferguson, Baltimore, and Dallas Protests. However, a visual computing researcher quickly noticed a large spike in tweets around the Dallas protest that are for ALM (below the center line) \textit{“Oh, I see… Cops were killed in the protest. These people care for the cops (“blue lives”) who were killed.”} [WF 2]. 

Finally, the communications experts recalled that in the SAH analysis (first case study), Care was correlated with mask usage during COVID-19. Examining the counties expressing Care in this second study, they remarked on the shift in terms of geographical coverage: \textit{“Care [in this second study] and Care [in the first study] is [not] correlated. That is counter-intuitive”} [WF 1]. Our collaborators theorized that this may reflect a shift in moral sentiment in partially republican areas between 2016 and after the 2020 pandemic, and a shift in priorities of the GOP rhetoric towards more Libertarian Rhetoric and away from Care [WF 2].

\subsection{Expert Feedback}
Overall, the domain experts found the MOTIV interface \textit{“valuable on multiple fronts”}. In the data collection stage, they stated it helped them \textit{“identify errors in the filtering process and refine queries”}. In the exploratory data analysis stage, it helped them \textit{“understand the relationships between stance, moral frames, sentiment, and political affiliation”}. Through the map interface, the group stated they were \textit{“able to identify some of the demographic biases in twitter opinions”}.
In later stages feedback was more explicit and supportive \textit{“Oh, wow, the bubbles, that's *powerful*. I have to say, I am very impressed with the work. I am blown away by what you do. That's a really powerful graphic.”}.
Finally, in the hypothesis generation stage, the interface helped the group \textit{“find counterintuitive findings first (e.g., no moral frame was associated with just support or just opposition to SAH attitudes) which we looked into further”}.

\begin{figure}[htb]
\begin{center}
    \includegraphics[width=\linewidth]{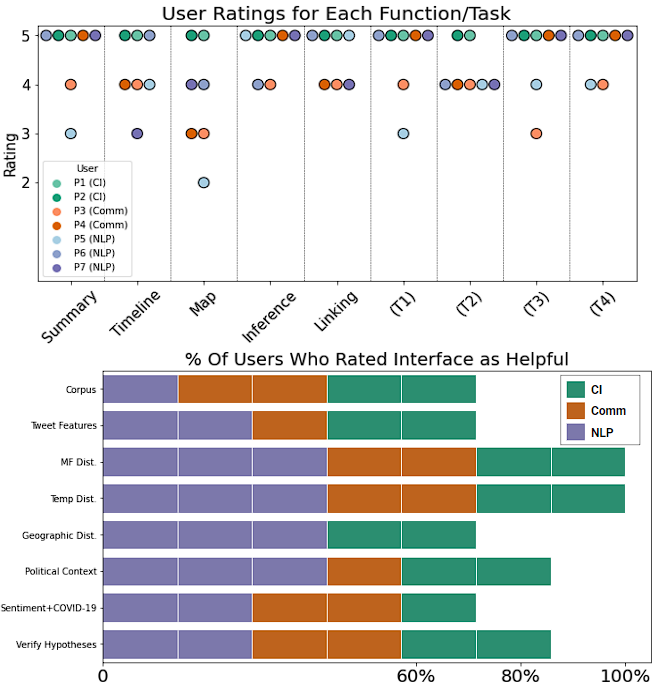}
    \caption{Results from user feedback questionnaire for users. (Top) Responses from the 5-point ratings for each component and analysis task of the system: T1-Popular frames; T2-Relevant tweet features; T3-Geopolitical or demographic trends; T4-Tweet trends over time. (Bottom) Responses from usefulness ratings of different tasks: understand Corpus; understand tweet features; understand Moral Frame, temporal, and geographic distributions;  understand political context of events;  correlate sentiment and COVID-19 rates; and verify new hypothesis about the data, respectively.}
    \label{fig:feedback1}
    \end{center}
\end{figure}

We asked the seven experts in NLP, Communications, and Causal Inference
(CI) to rate the perceived usefulness of each component of MOTIV on a 5-point scale, as well as its helpfulness in Identifying certain features: (T1) Popular frames. (T2) Relevant tweet features. (T3) Geopolitical/demographic trends. (T4) Tweet trends over time. In addition, we asked experts to rate specific system capabilities and helpful or not, and open-ended feedback.Results and the respondent’s domain expertise are shown in \cref{fig:feedback1}. The encodings were well received, while two users asked for a diagram explaining the map glyphs, which was later added into the system.

\section{Discussion and Conclusion}
MOTIV was developed through participatory design over the course of a developing project.
This project was built around a rapidly evolving study, in which the aims and data were constantly shifting during the entire course of the project. Despite these challenges, MOTIV's adoption as a research instrument by our collaborators is strong evidence of its value. The case studies and feedback we report further demonstrate that our approach, which blends data visualization, XAI, and social science, provides rich insights.

We thus discuss insights obtained from designing for these exceptional circumstances.

\textbf{Collaborative Design During Data Foraging}
Our project was started at the height of the COVID-19 pandemic, and included multiple collaborators for different domains, with disparate design goals and an uncertain dataset. As a result, we found many challenges during the concurrent design and data foraging stage. For example, we found collaborators expressed interest initially in an individual's political affiliation and how geographic location affected the spread of individual viral tweets. However, our data collection process yielded difficulties with linking tweets and retweets between individuals, which made looking at geographic spread impossible, and made inferring political affiliation at an individual level difficult. Additionally, challenges in automatic annotation of relevant tweets and moral frames meant our final dataset was much smaller than originally thought, which required significant refinement to the final design. Additionally, our original topic changed so quickly that newly acquired data was considered "obsolete" (e.g. Twitter's free API is now no longer available). As a result, we saw significant benefit from making the final design highly flexible in terms of the design and user control for exploration of different variables, which made adapting the interface to other problems, such as BLM, useful.

In terms of collaboration, we found that our collaborators had unclear or unrelated design goals for the dataset. Initially, collaborators stated they were mainly interested in a basic COVID-19 dashboard alongside a county map of political affiliation. However, analyzing implied goals and workflows during lab meetings and discussions led to different directions. As a result, careful note taking, and rapid development prototypes that used the actual data during the foraging stage in order to gain as much natural feedback as possible turned out to be a more effective way of soliciting design requirements.

Finally, when publishing results, the project met difficulties due to differing expectations: our collaborators and program officers expected valuable insights and a useful system, which the project provided, while reviewers alternatively anticipated large-scale corpuses, general or automated tools, or detailed validation of published NLP algorithms.

\textbf{Visual Complexity}
During our design process, a common issue we faced was reconciling the visual complexity necessitated by our task requirements with the simplicity required by our collaborators. Earlier iterations were often complex, as our original tasks were focused on inspecting many variables, with little insight into which were most important. We found that an approach grounded in visual scaffolding~\cite{marai2015visual} was most effective, where we built on encodings such as bar charts and choropleth maps that are familiar to users, and we added features and variations to address tasks that could not be met using traditional methods.

One example was the design of the glyph map. During prototyping collaborators showed a strong preference for choropleth maps due to their familiarity with it over unfamiliar methods that could show multiple variables. The glyph-based encoding helped in this sense, and also helped with the identification of major cities and counties of interest. By inspecting the alternative choropleth map, the viewers were able to mentally anchor the location of different areas in the glyph map even after being distorted by the force-layout, yielding positive feedback.

Ultimately, we found our collaborators’ visual literacy improved during the project. A focus on identifying well-motivated incremental changes to existing designs may help applications be more accessible and generalizable to domain experts and wider audiences. 

\textbf{Collaborative Hypothesis Testing}
A recurring challenge was to support the exploration of a large problem space to help identify interesting avenues for further investigation by collaborators. During the initial stages of our project, we found that a common workflow was that collaborators with computing backgrounds would share preliminary, exploratory data analysis using a variety of NLP techniques. Findings would be shared with experts in communications, who would identify potentially interesting findings. Follow-up statistical testing could then be performed to identify useful results.

However, most user-centered design approaches focus on the workflows of individuals, which can generally be obtained based on input from the user themselves. We found benefit in characterizing the workflows that occur through interactions between domain experts across domains, and focusing on adding in visualization that helps support gaps in information sharing between groups. For example, the Inference panel helped share results between the workflows of our statistical researchers and those with backgrounds in communications and moral foundation theory.

\textbf{Assumptions and Limitations}
MOTIV inherits the biases in the data we use. For example, most Twitter users tend to be younger and more democratic than the average American~\cite{pewtwittersurvey}.
While we have only looked at relatively small (< 2000 tweets) datasets, we have explored the use of MOTIV for a significantly larger set of over 100,000 tweets annotated using a combination of machine learning approaches~\cite{elenaDSAA}. In terms of generalizability the summary, map, and inference views can scale to arbitrary sizes. However, the Timeline view would need adjustments, as it shows encodings for all tweets simultaneously. Based on prototyping  (see supplemental materials), we found that variations of aggregated timelines and sparklines work well for showing trends, while context and influential tweets could be investigated on-demand by showing the most popular tweets within user-selected criteria at different time points of interest. The usage of improved automatic labeling could allow the system to work with arbitrary datasets, although current state of the art models do not perform well compared to manual labeling~\cite{elenaDSAA, hoover2020moral}.

In this work we introduced a novel methodology for exploring moral frame political discourse via analysis of social media. Our approach draws on methods from data visualization, explainable machine learning, and social science to provide rich insights into how the public formulates arguments over social media. By integrating Moral Foundations theory with custom visual encodings and interactions, we provide a novel and rich approach to Twitter data visualization systems. Through a detailed analysis in two case studies of tweets related to Stay-at-home orders in the U.S. during the COVID-19 pandemic and the Black Lives Matter movement, we show this approach can identify key events that affect the nature of political discourse, even without the presence of explicit labels, in addition to insights into how moral values regarding politicized movements are disseminated by different social groups. We identify design lessons relevant to working with domain scientists who have limited visual literacy, yet are keenly interested in quick hypothesis generation and testing. While we focus on the application of a specific set of Moral Foundations to frame our analysis, our approach could also apply to problems centered around comparing classes of tweets, such as in topic modeling, or when clustering latent variables learned using recurrent neural networks.

\section*{Acknowledgments}
This work was partially supported by NSF award IIS-2031095 and NIH award NCI-R01 CA258827.



\printbibliography
\end{document}